\documentclass{emulateapj}
\usepackage{natbib}
\usepackage{graphicx}
\usepackage{epstopdf}
\usepackage{color}

\newcommand{\Ha}{H$\alpha$}

\newcommand{\oiii}{[OIII]$\lambda$5007}

\newcommand{\lya}{Ly$\alpha$}
\newcommand{\fesc}{$f^{Ly\alpha}_{esc}$}
\newcommand{\fwhm}{FWHM$_{m}$}
\def\arcsec{\hbox{$^{\prime\prime}$}}

\begin{document}
\title{\lya\ and UV Sizes of Green Pea Galaxies}

\author{Huan Yang\altaffilmark{1,2}, Sangeeta Malhotra\altaffilmark{2}, James E. Rhoads\altaffilmark{2}, Claus Leitherer\altaffilmark{3}, Aida Wofford\altaffilmark{4}, Tianxing Jiang\altaffilmark{2},  Junxian Wang\altaffilmark{1} }

\altaffiltext{1}{CAS Key Laboratory for Research in Galaxies and Cosmology, Department of Astronomy, University of Science and Technology of China;  huan.y@asu.edu}
\altaffiltext{2}{Arizona State University, School of Earth and Space Exploration}
\altaffiltext{3}{Space Telescope Science Institute}
\altaffiltext{4}{National Autonomous University of Mexico, Institute of Astronomy}

\begin{abstract}

Green Peas are nearby analogs of high-redshift \lya-emitting galaxies (LAEs). To probe their \lya\ escape, we study the spatial profiles of \lya\ and UV continuum emission of 24 Green Pea galaxies using the Cosmic Origins Spectrograph (COS) on Hubble Space Telescope (HST). We extract the spatial profiles of \lya\ emission from their 2D COS spectra, and of UV continuum from both the 2D spectra and NUV images. The \lya\ emission shows more extended spatial profiles than the UV continuum in most Green Peas. The deconvolved Full Width Half Maximum (FWHM) of the \lya\ spatial profile is about 2 to 4 times that of the UV continuum in most cases. 
Since Green Peas are analogs of high-$\it{z}$ LAEs, it suggests that most high-$\it{z}$ LAEs likely have larger \lya\ sizes than UV sizes.
We also compare the spatial profiles of \lya\ photons at blueshifted and redshifted velocities in eight Green Peas with sufficient data quality, and find the blue wing of the \lya\ line has a larger spatial extent than the red wing in four Green Peas with comparatively weak blue \lya\ line wings. 
We show that Green Peas and MUSE $z=3-6$ LAEs have similar \lya\ and UV continuum sizes, which probably suggests starbursts in both low-$z$ and high-$z$ LAEs drive similar gas outflows illuminated by \lya\ light.
Five Lyman continuum (LyC) leakers in this sample have similar \lya\ to UV continuum size ratios ($\sim1.4-4.3$)  to the other Green Peas, indicating their LyC emission escape through ionized holes in the interstellar medium.
\end{abstract}

\section{Introduction}
The \lya\ emission line is a key tool in discovering and studying high redshift galaxies (e.g. Dey et al. 1998; Hu et al. 1998; Rhoads et al. 2000; Ouchi et al. 2003; Matthee et al. 2014; Zheng et al. 2016).  At $z>6$, the \lya\ luminosity, \lya\ equivalent width (EW), and spatial clustering of \lya\ emitting galaxies (LAEs) are important probes of the reionization of Universe (e.g. Malhotra \& Rhoads 2004; Kashikawa et al. 2011; Treu et al. 2012; Pentericci et al. 2014; Tilvi et al. 2014). To understand LAEs and reionization requires us to understand how \lya\ escape from galaxies. 

Since \lya\ is a resonant line, the \lya\ escape depends on the amount of dust, the HI gas column density ($N_{HI}$), the velocity distribution of HI gas, and the geometric distribution of HI gas and dust (e.g. Neufeld 1990; Charlot \& Fall 1993; Verhamme et al. 2006; Dijkstra et al. 2006). One important indicator of \lya\ escape processes is the \lya\ spatial distribution. 
The \lya\ emission would be confined to HII regions and have similar size to the UV continuum emission  if most \lya\ photons escape from ionized holes in the interstellar medium (ISM). Instead, if most \lya\ photons diffuse out of galaxy through numerous resonant scatterings, the \lya\ emission would be more extended than the UV continuum (e.g. \"{O}stlin et al. 2009; Zheng et al. 2010; Hayes et al. 2014). 

Prior HST studies of \lya\ morphology in low redshift starburst galaxies usually show diffuse \lya\ emission in the outer part of galaxy and sometimes \lya\ absorption in the center of galaxy (Kunth et al. 2003; Mas-Hesse et al. 2003; \"{O}stlin et al. 2009, Hayes et al. 2005, 2014). But most of those low redshift starbursts have much lower \lya\ EW ($EW<20$ \AA) and \lya\ escape fraction (\fesc) than high-$\it{z}$ LAEs. Since \lya\ photons escape more easily and probably have fewer scatterings in high-$\it{z}$ LAEs, it is reasonable to suppose that LAEs with high \lya\ EW may have compact \lya\ sizes. Due to the faintness of high-$\it{z}$ LAEs, there are only two studies of \lya\ size with high resolution HST narrow-band imaging for a few high-$\it{z}$ LAEs (Bond et al. 2010; Finkelstein et al. 2011), and they reached contradictory conclusions: Bond et al. (2010) suggested \lya\ sizes are compact and similar to UV continuum emission; but Finkelstein et al. (2011) suggested \lya\ appears larger than the UV continuum. 

Many ground based studies of \lya\ morphology suggest that a large scale faint \lya\ halo is common in high-$\it{z}$ \lya\ galaxies due to the scatterings of \lya\ photons by the HI gas in circum-galactic medium (e.g. Moller $\&$ Warren 1998; Swinbank et al. 2007; Rauch et al. 2008;  Steidel et al. 2011; Mastuda et al. 2012;  Feldmeier et al. 2013; Momose et al. 2015;  Wisotzki et al. 2015; Matthee et al. 2016). As the ground based data has low spatial resolution, however, it is still unclear if the \lya\ morphology of LAEs on galactic scales is compact or larger than the UV continuum, and if they show central \lya\ absorption.

Green Pea galaxies are compact starburst galaxies with strong \oiii\ emission lines (EW(\oiii)$>300$ \AA) in the nearby universe (Cardamone et al. 2009). They have strong \lya\ emission lines (Jaskot et al. 2014; Henry et al. 2015; Yang et al. 2016); and their \lya\ EW distribution is similar to high-$\it{z}$ LAEs (Yang et al. 2016). Five Green Peas in our sample also show Lyman continuum emission (Izotov et al. 2016). In this paper, we study the spatial distribution of \lya\ and UV emission of 24 Green Peas with HST-COS, compare the spatial profiles of \lya\ photons at blue and red velocities, and discuss the implications to \lya\ and LyC escape.

\begin{figure*}[!ht]
  \centering
    \includegraphics[width=0.86\textwidth]{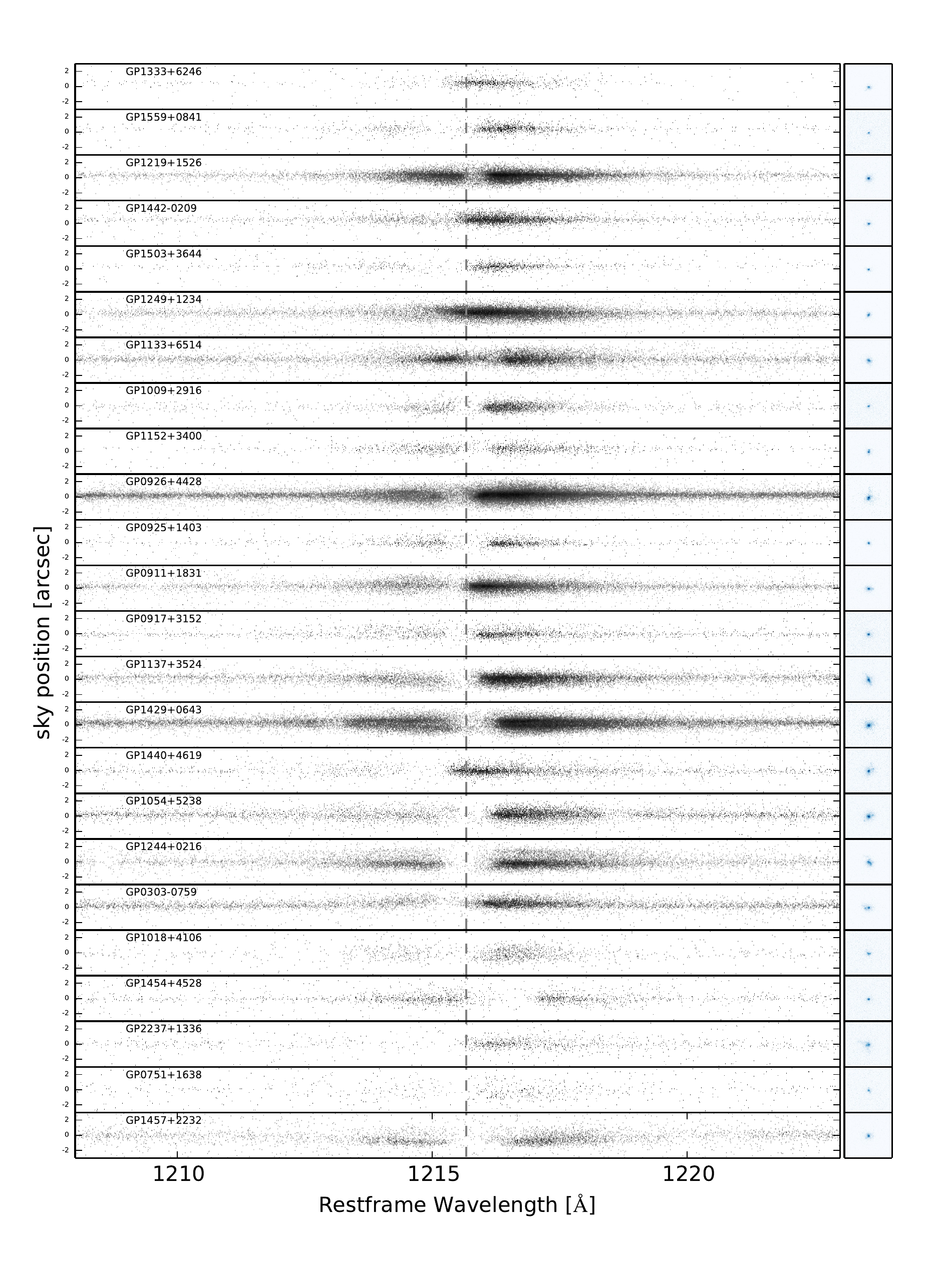}
 \caption{The 2D FUV spectra and NUV images of these 24 Green Peas. In the 2D spectra, X axis is along the dispersion direction and Y axis is along the sky direction. The COS aperture is a 2.5 arc-second diameter circle. The dashed vertical line marks the restframe wavelength of \lya. The NUV images (6\arcsec$\times$6\arcsec) are at the same orientation as the 2D spectra. All NUV images have the same range of color-bar in log-scale. These 24 galaxies are sorted by decreasing \fesc\ from top to bottom. The ID of each galaxy is marked in each panel. }
\end{figure*}

\begin{figure*}[!ht]
\centering
  \includegraphics[width=1.03\textwidth]{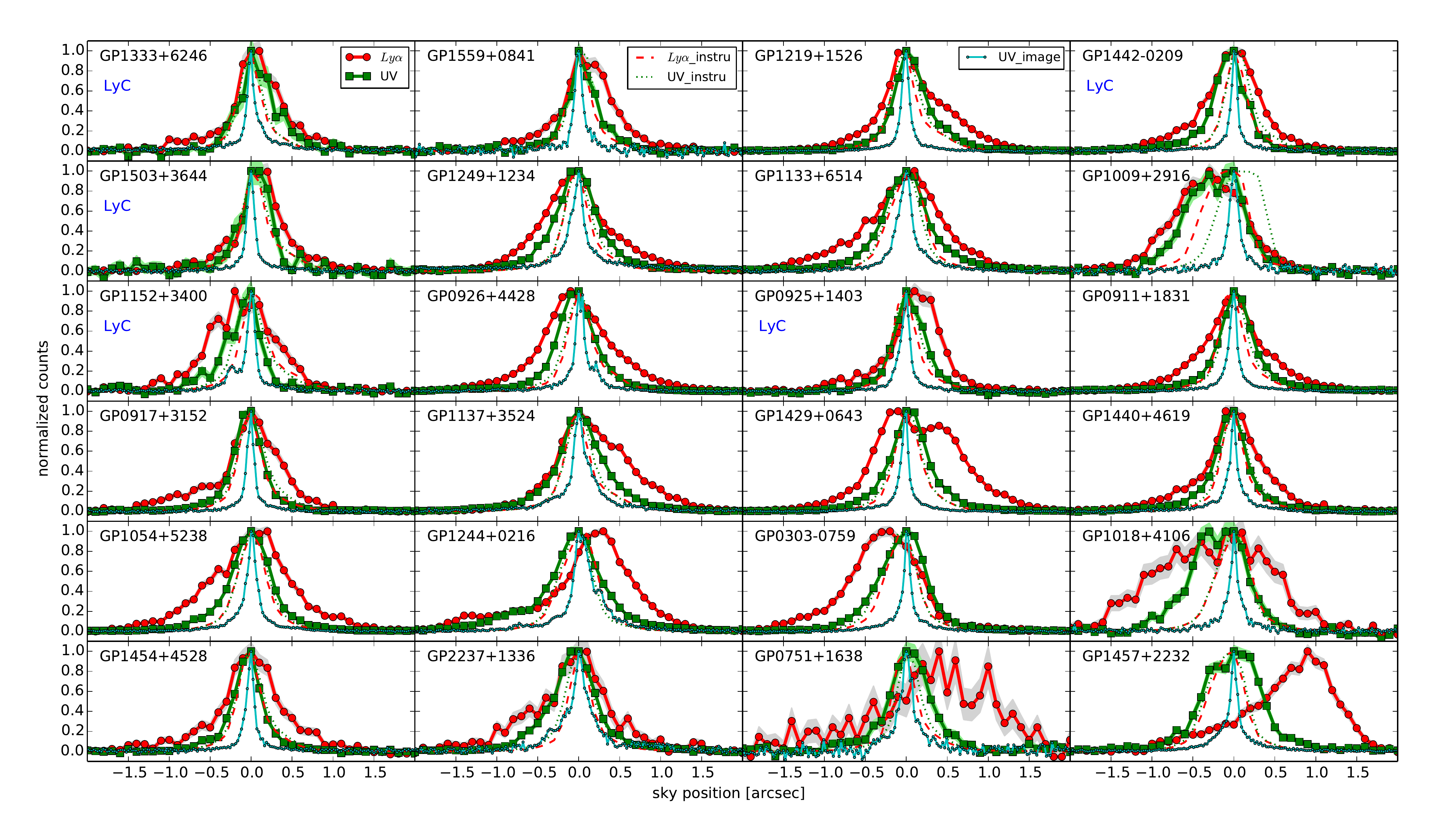}
 \caption{The normalized spatial profiles along the sky direction for the 24 Green Peas. In each panel, the solid green line with square marker shows the spatial profiles of the UV continuum emission measured from COS, and the solid red line with circle marker shows the spatial profiles of total \lya\ emission. The shaded grey (light-green) regions of the solid red (green) lines show the 1$\sigma$ errors of the \lya\ (UV) spatial profiles. The dotted green and dashed red lines show the instrumental spatial profiles for UV continuum and \lya\ emission respectively. The instrumental spatial profiles are derived from observations of a white dwarf point source (Section 2). The solid cyan line with dot marker shows the spatial profiles of UV continuum emission measured from NUV acquisition image along the spatial direction of 2D spectra. These 24 galaxies are sorted by decreasing \fesc\ from top to bottom and from left to right. Five LyC leakers in this sample (Izotov et al. 2016) are marked with ``LyC". 
}
\end{figure*}

\begin{deluxetable*}{lcccccccc}
\tabletypesize{\scriptsize}
\centering
\tablewidth{0pt}
\tablehead{
\colhead{ID} & \colhead{RA} & \colhead{DEC} & \colhead{Redshift} & \colhead{EW(\lya)} & \colhead{\fesc} & \colhead{WP} & \colhead{LP\#} & \colhead{GO\#} \\ 
\colhead{} & \colhead{} & \colhead{}  & \colhead{} & \colhead{\AA} & \colhead{} & \colhead{\AA} & \colhead{} & \colhead{} \\
\colhead{(1)} & \colhead{(2)} & \colhead{(3)} & \colhead{(4)} & \colhead{(5)} & \colhead{(6)} & \colhead{(7)} & \colhead{(8)} & \colhead{(9)} 
} 
\startdata
GP1333+6246$^{a}$ & 13:33:03.94 & +62:46:03.7 & 0.318124 & 65.3 & 1.066 & 1623 & 3 & 13744 \\
GP1559+0841 & 15:59:25.98 & +08:41:19.1 & 0.297036 & 89.0 & 0.682 & 1623 & 3 & 14201 \\
GP1219+1526 & 12:19:03.98 & +15:26:08.5 & 0.195599 & 157.5 & 0.672 & 1623 & 2 & 12928 \\
GP1442$-$0209$^{a}$ & 14:42:31.37 & $-$02:09:52.8 & 0.293669 & 127.9 & 0.408 & 1623 & 3 & 13744 \\
GP1503+3644$^{a}$ & 15:03:42.82 & +36:44:50.8 & 0.355689 & 99.6 & 0.402 & 1623 & 3 & 13744 \\
GP1249+1234 & 12:48:34.64 & +12:34:02.9 & 0.263389 & 94.8 & 0.384 & 1623 & 2 & 12928 \\
GP1133+6514 & 11:33:03.80 & +65:13:41.3 & 0.241397 & 35.3 & 0.352 & 1600 & 2 & 12928 \\
GP1009+2916 & 10:09:18.99 & +29:16:21.5 & 0.221918 & 62.5 & 0.335 & 1589 & 3 & 14201 \\
GP1152+3400$^{a}$ & 11:52:04.88 & +34:00:49.9 & 0.341946 & 67.5 & 0.260 & 1623 & 3 & 13744 \\
GP0926+4428 & 09:26:00.44 & +44:27:36.5 & 0.180690 & 40.8 & 0.245 & 1611 & 1 & 11727 \\
GP0925+1403$^{a}$ & 09:25:32.37 & +14:03:13.1  & 0.301211 & 83.0 & 0.171 & 1623 & 3 & 13744 \\
GP0911+1831 & 09:11:13.34 & +18:31:08.2 & 0.262200 & 49.5 & 0.155 & 1623 & 2 & 12928 \\
GP0917+3152 & 09:17:02.52 & +31:52:20.6 & 0.300364 & 31.0 & 0.138 & 1623 & 3 & 14201 \\
GP1137+3524 & 11:37:22.14 & +35:24:26.7 & 0.194390 & 33.4 & 0.130 & 1623 & 2 & 12928 \\
GP1429+0643 & 14:29:47.03 & +06:43:34.9 & 0.173509 & 35.7 & 0.103 & 1600 & 2 & 13017 \\
GP1440+4619 & 14:40:09.94 & +46:19:36.9 & 0.300758 & 26.8 & 0.101 & 1623 & 3 & 14201 \\
GP1054+5238 & 10:53:30.83 & +52:37:52.9 & 0.252638 & 10.7 & 0.068 & 1611 & 2 & 12928 \\
GP1244+0216 & 12:44:23.37 & +02:15:40.4 & 0.239426 & 40.0 & 0.065 & 1600 & 2 & 12928 \\
GP0303$-$0759 & 03:03:21.41 & $-$07:59:23.2 & 0.164880 & 7.2 & 0.050 & 1589 & 2 & 12928 \\
GP1018+4106 & 10:18:03.24 & +41:06:21.1 & 0.237052 & 26.1 & 0.047 & 1600 & 3 & 14201 \\
GP1454+4528 & 14:54:35.58 & +45:28:56.3 & 0.268505 & 23.0 & 0.047 & 1623 & 3 & 14201 \\
GP2237+1336 & 22:37:35.06 & +13:36:47.0 & 0.293501 & 9.9 & 0.034 & 1623 & 3 & 14201 \\
GP0751+1638 & 07:51:57.80 & +16:38:13.2 & 0.264713 & 8.8 & 0.024 & 1623 & 3 & 14201 \\
GP1457+2232 & 14:57:35.13 & +22:32:01.8 & 0.148611 & 5.3 & 0.010 & 1577 & 2 & 13293 
\enddata
\tablecomments{Column descriptions: (5-6) restframe \lya\ equivalent width, and \lya\ escape fraction from Yang et al. (2016b, in-prep); (7) Central wavelength position of G160M grating; (8) COS lifetime position (LP); (9) HST programs: GO14201 (PI S. Malhotra), GO13744 (PI T. Thuan; Izotov et al. 2016), GO13293 (PI A. Jaskot; Jaskot et al. 2014), GO12928 (PI A. Henry; Henry et al. 2015), GO11727 and GO13017 (PI T. Heckman; Heckman et al. 2011; Alexandroff et al. 2015). } 
\tablenotetext{a}{These are confirmed LyC leakers from Izotov et al. (2016).}
\end{deluxetable*}

\begin{deluxetable*}{lcccccc}
\tabletypesize{\scriptsize}
\centering
\tablewidth{0pt}
\tablehead{
\colhead{ID}  & \colhead{FWHM$_{m}$(\lya)} & \colhead{FWHM$_{m}$(FUV)} &  \colhead{$FWHM_{d}(Ly\alpha)$} & \colhead{$FWHM_{d}(FUV)$} & \colhead{FWHM(NUV)} & \colhead{$\frac{FWHM(Ly\alpha)}{FWHM(UV)}$}  \\ 
\colhead{} & \colhead{arcsec} & \colhead{arcsec} & \colhead{arcsec} & \colhead{arcsec} & \colhead{arcsec} & \colhead{} \\
\colhead{(1)} & \colhead{(2)} & \colhead{(3)} & \colhead{(4)} & \colhead{(5)} & \colhead{(6)} & \colhead{(7)} 
} 
\startdata
GP1333+6246$^{a}$ & 0.56$\pm$0.04 & 0.38$\pm$0.08 & 0.24$^{+0.04}_{-0.03}$ & 0.08$^{+0.07}_{-0.07}$ & 0.09 & 2.62$^{+0.80}_{-0.52}$ \\
GP1559+0841 & 0.56$\pm$0.04 & 0.35$\pm$0.04 & 0.25$^{+0.04}_{-0.04}$ & 0.04$^{+0.04}_{-0.03}$ & 0.11 & 2.27$^{+0.67}_{-0.55}$ \\
GP1219+1526 & 0.65$\pm$0.03 & 0.42$\pm$0.01 & 0.29$^{+0.01}_{-0.03}$ & 0.06$^{+0.01}_{-0.01}$ & 0.11 & 2.65$^{+0.43}_{-0.47}$ \\
GP1442$-$0209$^{a}$ & 0.70$\pm$0.04 & 0.44$\pm$0.03 & 0.37$^{+0.04}_{-0.04}$ & 0.12$^{+0.03}_{-0.03}$ & 0.08 & 3.00$^{+1.29}_{-0.82}$ \\
GP1503+3644$^{a}$ & 0.48$\pm$0.04 & 0.38$\pm$0.06 & 0.15$^{+0.03}_{-0.03}$ & 0.06$^{+0.06}_{-0.04}$ & 0.11 & 1.39$^{+0.43}_{-0.36}$ \\
GP1249+1234 & 0.75$\pm$0.02 & 0.56$\pm$0.02 & 0.39$^{+0.03}_{-0.01}$ & 0.19$^{+0.01}_{-0.01}$ & 0.17 & 2.00$^{+0.31}_{-0.20}$ \\
GP1133+6514 & 0.96$\pm$0.06 & 0.58$\pm$0.01 & 0.54$^{+0.07}_{-0.07}$ & 0.17$^{+0.01}_{-0.01}$ & 0.20 & 2.70$^{+0.69}_{-0.56}$ \\
GP1009+2916 & 0.95$\pm$0.06 & 0.82$\pm$0.04 & 0.46$^{+0.07}_{-0.07}$ & 0.30$^{+0.06}_{-0.06}$ & 0.14 & 1.50$^{+0.61}_{-0.42}$ \\
GP1152+3400$^{a}$ & 0.87$\pm$0.09 & 0.44$\pm$0.07 & 0.47$^{+0.08}_{-0.10}$ & 0.08$^{+0.06}_{-0.07}$ & 0.11 & 4.28$^{+1.32}_{-1.19}$ \\
GP0926+4428 & 0.82$\pm$0.01 & 0.49$\pm$0.01 & 0.44$^{+0.01}_{-0.01}$ & 0.14$^{+0.01}_{-0.01}$ & 0.12 & 3.20$^{+0.47}_{-0.38}$ \\
GP0925+1403$^{a}$ & 0.58$\pm$0.04 & 0.42$\pm$0.03 & 0.26$^{+0.03}_{-0.03}$ & 0.08$^{+0.01}_{-0.03}$ & 0.12 & 2.19$^{+0.50}_{-0.41}$ \\
GP0911+1831 & 0.62$\pm$0.02 & 0.44$\pm$0.01 & 0.28$^{+0.03}_{-0.01}$ & 0.11$^{+0.01}_{-0.01}$ & 0.10 & 2.50$^{+0.64}_{-0.39}$ \\
GP0917+3152 & 0.62$\pm$0.04 & 0.39$\pm$0.01 & 0.30$^{+0.04}_{-0.04}$ & 0.04$^{+0.01}_{-0.03}$ & 0.10 & 3.05$^{+0.80}_{-0.66}$ \\
GP1137+3524 & 0.88$\pm$0.02 & 0.56$\pm$0.01 & 0.49$^{+0.03}_{-0.01}$ & 0.19$^{+0.01}_{-0.01}$ & 0.18 & 2.50$^{+0.35}_{-0.23}$ \\
GP1429+0643 & 1.18$\pm$0.02 & 0.49$\pm$0.01 & 0.90$^{+0.01}_{-0.07}$ & 0.12$^{+0.01}_{-0.01}$ & 0.09 & 7.22$^{+1.03}_{-1.22}$ \\
GP1440+4619 & 0.62$\pm$0.04 & 0.42$\pm$0.01 & 0.29$^{+0.04}_{-0.03}$ & 0.07$^{+0.01}_{-0.01}$ & 0.08 & 3.64$^{+0.98}_{-0.65}$ \\
GP1054+5238 & 0.99$\pm$0.07 & 0.54$\pm$0.01 & 0.62$^{+0.10}_{-0.08}$ & 0.17$^{+0.01}_{-0.01}$ & 0.13 & 3.75$^{+0.98}_{-0.75}$ \\
GP1244+0216 & 0.82$\pm$0.03 & 0.63$\pm$0.02 & 0.39$^{+0.04}_{-0.03}$ & 0.22$^{+0.01}_{-0.03}$ & 0.24 & 1.62$^{+0.37}_{-0.25}$ \\
GP0303$-$0759 & 0.94$\pm$0.04 & 0.63$\pm$0.02 & 0.55$^{+0.04}_{-0.04}$ & 0.19$^{+0.01}_{-0.01}$ & 0.08 & 2.86$^{+0.45}_{-0.39}$ \\
GP1018+4106 & 1.75$\pm$0.27 & 0.76$\pm$0.04 & 2.15$^{+1.80}_{-0.83}$ & 0.33$^{+0.04}_{-0.03}$ & 0.10 & 6.46$^{+6.50}_{-2.94}$ \\
GP1454+4528 & 0.62$\pm$0.05 & 0.37$\pm$0.02 & 0.29$^{+0.04}_{-0.06}$ & 0.01$^{+0.04}_{-0.01}$ & 0.10 & 2.91$^{+0.79}_{-0.77}$ \\
GP2237+1336 & 0.68$\pm$0.11 & 0.56$\pm$0.03 & 0.36$^{+0.10}_{-0.11}$ & 0.18$^{+0.03}_{-0.01}$ & 0.20 & 1.80$^{+0.74}_{-0.67}$ \\
GP0751+1638 & 0.97$\pm$0.32 & 0.53$\pm$0.04 & 0.62$^{+0.42}_{-0.32}$ & 0.19$^{+0.04}_{-0.03}$ & 0.17 & 3.21$^{+3.04}_{-1.92}$ \\
GP1457+2232 & 0.92$\pm$0.09 & 0.80$\pm$0.03 & 0.50$^{+0.10}_{-0.10}$ & 0.33$^{+0.04}_{-0.03}$ & 0.10 & 1.50$^{+0.45}_{-0.43}$
\enddata
\tablecomments{Column descriptions: (2-3) measured Full Width Half Maximum (FWHM) of \lya\ and UV spatial profiles.  (4-5) $FWHM_{d}(Ly\alpha)$ and $FWHM_{d}(FUV)$ are deconvolved FWHM of \lya\ and UV derived by mapping measured FWHM to intrinsic values (see section 3.1 and figure 3). (6) FWHM of 1D NUV profile. We convert the 2D NUV images into a 1D profile along the sky direction of spectra.  (7) Ratios of FWHM$(Ly\alpha)$ to FWHM(UV). When calculating the ratio, we use the larger one of $FWHM_{d}(FUV)$ and FWHM(NUV). These 24 galaxies are sorted by decreasing \fesc\ from top to bottom.} 
\tablenotetext{a}{These are confirmed LyC leakers from Izotov et al. (2016).}
\end{deluxetable*}

\section{Observations and Data Analysis}
In Yang et al. (2017), we assemble a sample of 43 Green Peas with HST-COS spectroscopic observations. Comparing to the parent sample of Green Peas in Cardamone et al. (2009), this sample covers the full ranges of properties, such as dust extinction, metallicity, and star formation rate (figure 1 in Yang et al. 2017). Thus it is a representative sample of Green Peas.
From this sample, we select 24 Green Peas which have good spatial resolution (Full Width at Half Maximum, FWHM$\sim0.3-0.4\arcsec$ for point source) in their 2D spectra. Since the COS FUV channel is not corrected for spherical aberration, the cross-dispersion resolution of COS FUV spectra depends on the chosen grating, the wavelength position (WP) of the grating, and the wavelength (COS ISR2013\_07). The grating and WP are chosen based on considerations of wavelength coverages and the gap in FUV detectors, thus varies mostly with the redshifts. Although this sample only covers a small redshift range ($\sim0.1-0.3$), a slightly different redshift, thus a different grating WP, can result in very different spatial resolution in the 2D spectra. So these 24 selected Green Peas are not statistically different from the sample of 43 Green Peas in obvious ways.

High resolution NUV acquisition images were taken with the COS acquisition mode  ACS/IMAGE for all 24 Green Peas. Their FUV spectra were taken with the 2.5\arcsec\ diameter Primary Science Aperture and the G160M grating, which has the best spatial resolution in all COS gratings. 

The COS FUV grating G160M has five WP -- 1577\AA, 1589\AA, 1600\AA, 1611\AA, 1623\AA. The WP=1623\AA\ has the best spatial resolution and 15/24 of Green Peas are taken in this WP.  The COS spatial resolutions are about $0.3-0.4$\arcsec\ for point source and stable with wavelength for the WP=1600\AA, 1611\AA, and 1623\AA, but are larger and vary moderately with wavelength for the WP=1577\AA\ and 1589\AA. We generally avoid using objects with WP=1577\AA\ or 1589\AA\ except for three cases where their \lya\ emission lines are in wavelength ranges with small spatial resolution. The WP of each object is shown in Table 1. 
  
We retrieved COS spectra of these 24 Green Peas from the HST MAST archive after they have been processed through the standard COS pipeline. The calibrated two dimensional \lya\ and FUV spectra are shown in figure 1. We extract the spatial profiles of \lya\ along the sky direction by summing the spectra in a wavelength range about 1211$-$1220 \AA\ along the dispersion direction.  We extract the spatial profiles of FUV continuum in wide wavelength ranges of a few tens Angstroms near \lya\ lines in the same spectra segment. 
Then we sum the spatial profiles from spectra taken at different central wavelengths or FP-POS settings for each Green Pea. 
In figure 2, we show their normalized spatial profiles of \lya\ and FUV continuum light. The pixel scale along the sky direction is 0.1\arcsec/pixel. Since the COS FUV detector counts photons, we assume the photon counts in each spatial bin follows Poisson statistics, and calculate its statistical error as counts$_{err}$ = (counts)$^{1/2}$. 

In figure 2, we also show the instrumental spatial profile of each object derived from observations of a point source in the same grating and WP (WD1057$+$719, CAL/COS 12806, PI: Derck Massa). Since the spatial resolution slightly varies with wavelength, the instrumental profiles are extracted for \lya\ and FUV continuum separately in the corresponding wavelength ranges that are used to extract the \lya\ and FUV spectra of each object. 

The response of COS/FUV detector decreases with usage, a process called gain-sag. To mitigate these gain-sag effects, COS/FUV spectra are moved to pristine locations of the detector, i.e. different lifetime positions (LP) every 2-3 years. Our sample spans on all three lifetime positions (LP1, LP2, and LP3). As we only use the data with small spatial resolution, the spatial profiles are separated from the insensitive detector regions of earlier LPs. The LP of each object is shown in Table 1. 

We then measure the \lya\ EW and \lya\ escape fraction (\fesc) of this sample (details in Yang et al. 2016b). The \fesc\ is defined as the ratio of the measured \lya\ flux to intrinsic \lya\ flux. Assuming case-B recombination, the intrinsic \lya\ flux is about 8.7 times dust extinction corrected \Ha\ flux measured from SDSS spectra. Thus the \fesc\ is \lya(observed)/(8.7$\times H\alpha_{corrected}$). In Table 1, we show their redshifts, \lya\ equivalent widths, and \lya\ escape fractions.

\begin{figure}[ht]
\centering
  \includegraphics[width=0.5\textwidth]{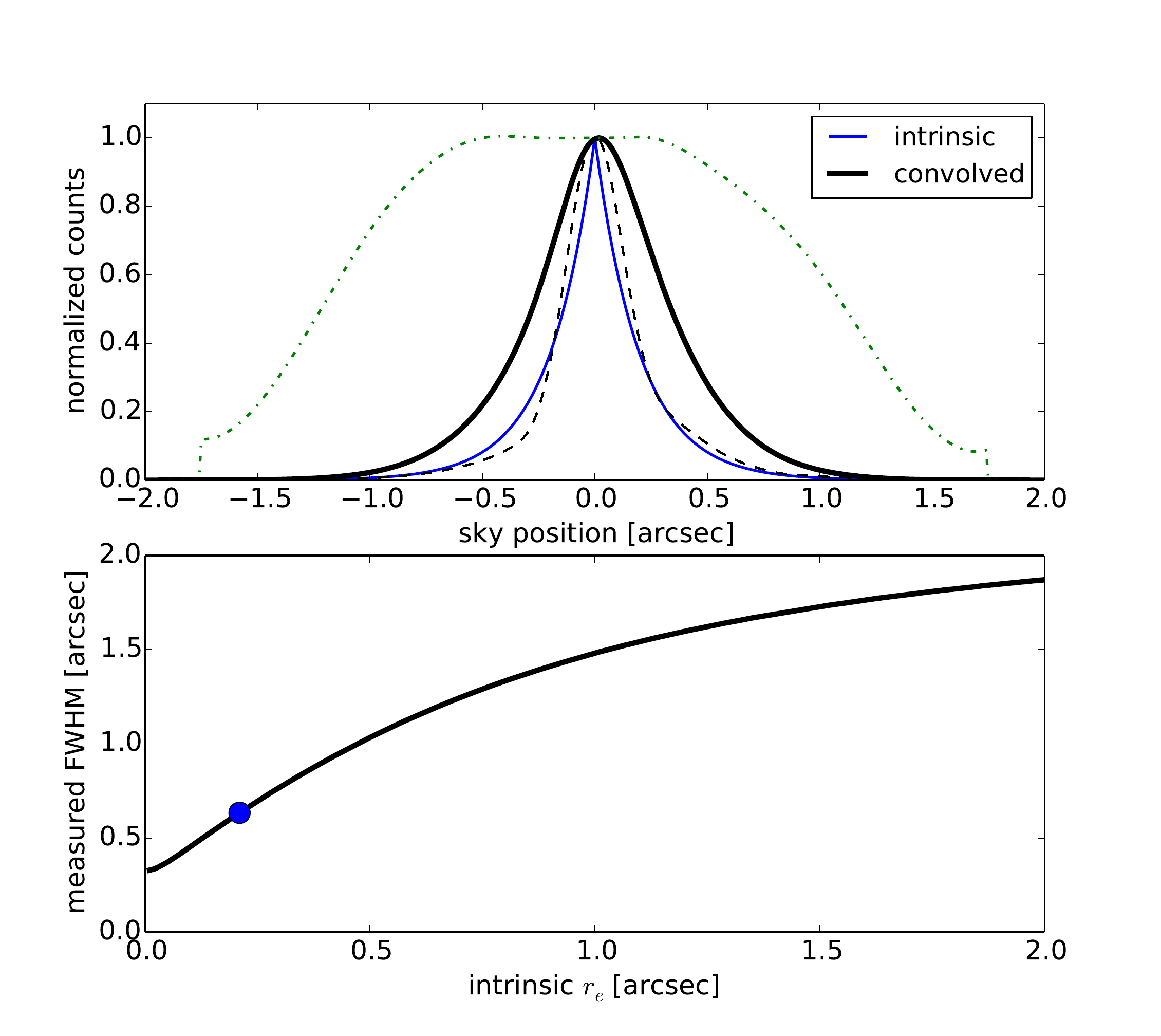}
 \caption{Top panel:  an example of the profile convolution. The dashed black line is a typical instrumental profile for \lya\ emission. The thin solid blue line shows the intrinsic spatial profile of \lya\ emission which is assumed to be an exponential profile (a typical profile with $r_{e}=0.2$\arcsec\ is shown).  The thick solid black line is the convolved profile after convolving the intrinsic profile with the instrumental profile and multiplying it by the throughput profile (dash-dotted green line). Bottom panel: the measured FWHM of the convolved profile as a function of the $r_{e}$ of the intrinsic profile. The blue point shows a typical value of the intrinsic \lya\ size in our sample. }
\end{figure}

\section{Compare Spatial Profiles of \lya\ and UV emission}

From the 2D spectra and 1D spatial profiles, we can see that the \lya\ emission comes from a larger region than the FUV emission in most of these 24 Green Peas. The spatial profiles of UV are only slightly larger than the instrumental profiles, but the spatial profiles of \lya\ are well resolved and show asymmetric spatial distributions in many cases.  In four cases with low \fesc\ (GP1457$+$2232, GP0303$-$0759, GP0752$+$1638, and GP1244$+$0216), \lya\ light shows a significant offset from the FUV continuum (similar to some high-$z$ LAEs in Micheva et al. 2015). In GP1429$+$0643, a large fraction of the \lya\ emission in the galactic center is absorbed, resulting in a double horned spatial profile.  

To characterize the size of spatial profile, we measure the FWHM (FWHM$_{m}$) of each profile. The FWHM is not sensitive to the depth of the observation. To get the error of FWHM$_{m}$ of each observed spatial profile, we simulate 1000 fake profiles by adding random Gaussian errors to the observed profile.  We measure the \fwhm\ of each fake profile and calculate the standard deviation of the 1000 fake profiles as the error of \fwhm\ for each observed spatial profile.   The measured \fwhm\ and its errors are shown in Table 2. We can see again that the \lya\ emission have significantly larger \fwhm\ than the UV continuum emission. 

\subsection{The Deconvolved Sizes of \lya\ and UV emission}

To estimate the deconvolved sizes, we assume the intrinsic \lya\ or UV emission follows an exponential profile with scale radius $r_{e}$ and convolve the exponential profile with the instrumental profile, so we get a relation between observed FWHM and intrinsic FWHM. Since the throughput begins to decrease when the offset from aperture center is larger than about 0.5\arcsec,  we multiply the convolved profile with a throughput curve of G160M retrieved from COS instrumental handbook. In figure 3, we show an example of the profile convolution and how the FWHM of convolved profile varies with the $r_{e}$ of intrinsic profile. We then calculate the deconvolved size of \lya\ emission as the FWHM of the exponential profile which has the same \fwhm\ as the observed \lya\ spatial profile. Since the measured \fwhm\ of \lya\ emission (about $0.6\arcsec-1.0\arcsec$) are within the angular ranges with $\gtrsim$ 80\% throughput, the \lya\ sizes are {\it not} underestimated due to attenuation at large offsets except in GP1018$+$4106 which has very large \lya\ size.

Since the NUV image has a spatial resolution of about 0.04\arcsec\ (less than 2 pixels at pixels scale of 0.0235\arcsec/pixel), the NUV emission of this sample are well resolved. We estimate the NUV size from the NUV acquisition image shown in figure 1 at the same orientation as the 2D spectra. We extract spatial profiles by summing the pixels in the image along the dispersion direction. Then we calculate the intrinsic NUV sizes as the FWHM of the NUV spatial profiles. The results are shown in Table 2.  

Ideally, the deconvolved FUV size and NUV size should be similar. However, when the observed FUV profile and instrumental profile are very similar, the deconvolution failed and resulted in very small deconvolved FUV size. To compare the sizes of \lya\ and UV emission, we use the larger one of $FWHM_{d}(FUV)$ and FWHM(NUV), so we get a conservative \lya\ to UV size ratio. The deconvolved \lya\ sizes are typically 2.6 times of the UV sizes and vary between 1.4 and 4.3 times for 22 out of the 24 Green Peas. In GP1429$+$0643 which has a double horned spatial profile, the deconvolved \lya\ FWHM is about 7 times the UV FWHM.  In GP1018$+$4106, the deconvolved \lya\ FWHM is badly constrained and can be $3.5-13$ times larger than the UV FWHM.

\begin{figure}[ht]
\centering
  \includegraphics[width=0.5\textwidth]{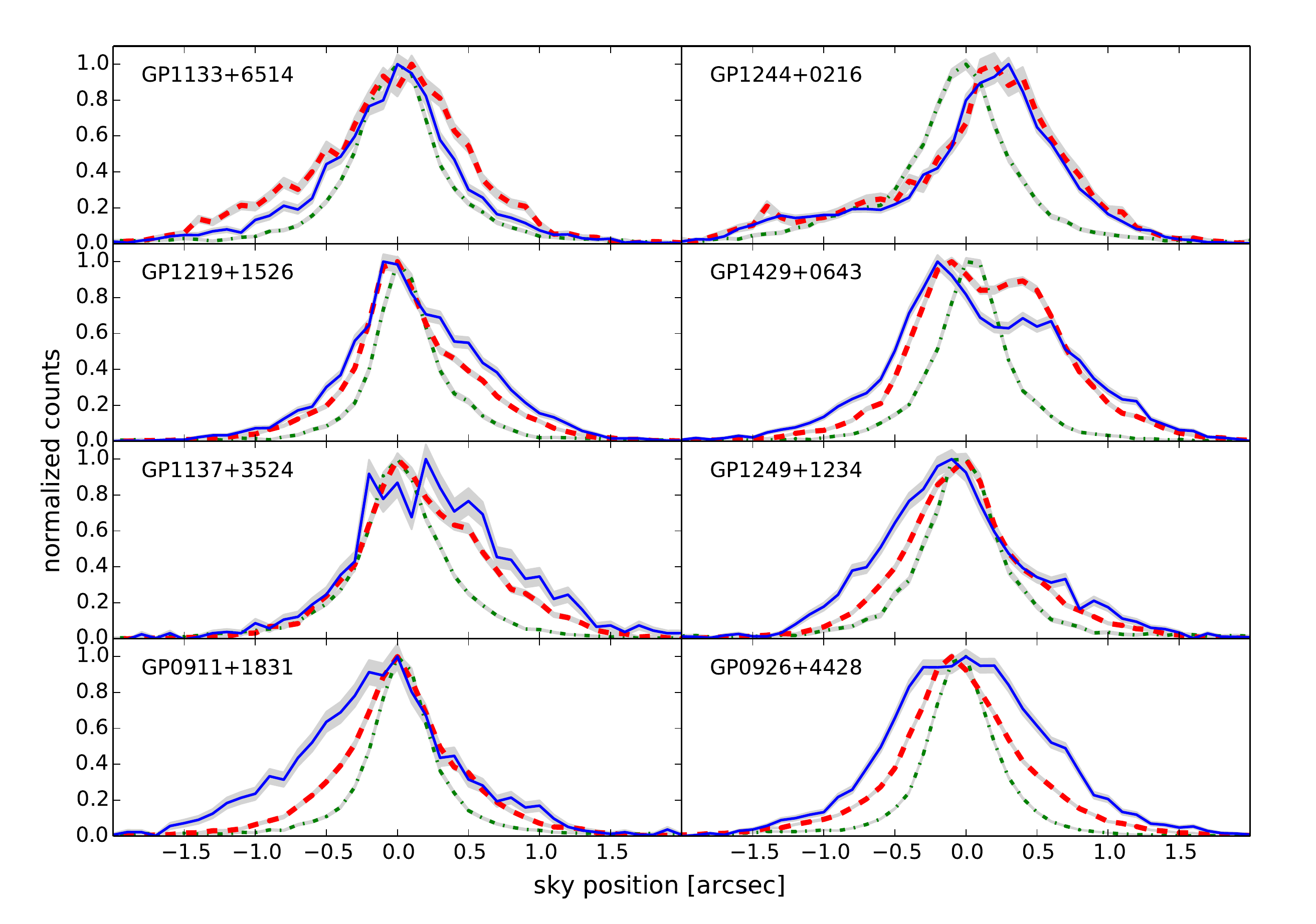}
 \caption{Comparison between the spatial profiles of the blueshifted and redshifted portions of the \lya\ emission lines for 8 objects with the best signal-to-noise ratio in the blue-part \lya\ emission.  Since the \lya\ velocity profiles are usually double-peaked, we define the blue-part (red-part) as the negative-velocity-side (positive-velocity-side) of the inter-peak dip of the \lya\ velocity profile. The solid blue lines, dashed red lines, and dotted green lines show the spatial profiles of blue-part \lya, red-part \lya, and UV continuum emission correspondingly. The shaded grey regions of lines show the 1$\sigma$ errors of the spatial profiles.}
\end{figure}

\section{Comparing Spatial Profiles of \lya\ Photons at Different Velocities}
Green Peas usually show double-peaked \lya\ velocity profiles (Jaskot et al. 2014; Henry et al. 2015; Yang et al. 2016).  The \lya\ photons with different velocities are scatterd differently by the HI gas. Since we have the 2D \lya\ spectra, we can compare the spatial profiles of \lya\ photons at different velocities. We define the blue-part (red-part) as the negative-velocity-side (positive-velocity-side) of the inter-peak dip of the \lya\ velocity profile. For one object (GP1249$+$1234) with a single peaked \lya\ velocity profile, we separate the blue-part and red-part by velocity=0. Then we extract the spatial profiles of the blue-part and red-part \lya\ emission. Since the blue-part is usually weaker than the red-part \lya\ emission, we show the 8 of 24 Green Peas with the best signal-to-noise ratio in the blue-part \lya\ emission. These 8 Green Peas also have relatively high \fesc. We compare their spatial profiles of blue-part and red-part \lya\ emission in figure 4. 

The spatial profiles of blue-part and red-part \lya\ emission are generally similar. But in four cases (GP1137$+$3524, GP1249$+$1234, GP0911$+$1831, and GP0926$+$4428), the blue-part \lya\ emission are more extended than the red-part \lya\ emission.  In the other four cases (GP1244$+$0216, GP1133$+$6514, GP1429$+$0643, and GP1219$+$1526), the blue-part and red-part \lya\ emission are very similar.  We also noticed that the \lya\ spatial profiles show a relation with the \lya\ velocity profiles -- the objects with weaker blue peak in \lya\ velocity profile (i.e. small flux ratio of blue-part to red-part \lya\ emission), such as GP1137$+$3524, GP0911$+$1831, and GP0926$+$4428, also have broader blue-part spatial profiles. On the other hand, GP1133$+$6514, which has the strongest blue peak in \lya\ velocity profile, seems to show slightly more compact blue-part \lya\ emission than the red-part \lya\ emission.  

In four Green Peas (GP1219$+$1526, GP1133$+$6514, GP0926$+$4428, and GP1429$+$0643), the \lya\ velocity profiles show large residual emission at velocity near zero. From their 2D spectra (figure 1), we find that the \lya\ emission at velocity near zero seems to have more extended \lya\ emission than the \lya\ emission at other velocities. 

Since the outflowing HI gas presented in many Green Peas has larger optical depth to the blue-part \lya\ photons than to the red-part \lya\ photons, we expect that the escaped blue-part \lya\ photons went through more scatterings on average and were scattered to larger radius. For the \lya\ photons at velocity near zero, the optical depth is the largest and their spatial profiles also show the largest sizes.

\section{Discussion}
\subsection{Comparison to Previous Results} 
Many studies measured the \lya\ morphology of some nearby star-forming galaxies with HST/STIS (Mas-Hesse et al. 2003) and HST/ACS images (e.g. Kunth et al. 2003, Hayes et al. 2005; \"{O}stlin et al. 2009, 2014). Mas-Hesse et al. (2003) analyzed the HST/STIS 2D spectra of \lya\ and UV emission and showed that both Haro 2 and IRAS 0833+6517 have  low \lya\ EW (6 \AA\ and 12 \AA) and larger \lya\ sizes than UV continuum sizes, and that their \lya\ peaks are offset from the peaks of UV continuum emission. 

The LARS program studies the \lya\ morphology of 14 nearby starburst galaxies (Hayes et al. 2014; \"{O}stlin et al. 2014). 9 out of the 14 galaxies have low \lya\ EW and escape fraction and they also show \lya\ absorption or weak \lya\ emission in the central part of galaxy and diffuse \lya\ emission in the outer part of the galaxy. The other 5 galaxies (LARS01, 02, 05, 07, and 14, LARS14 is the same galaxy GP0926$+$4428 in our sample) are LAEs with relatively high \lya\ EW and are comparable to most of the Green Peas in our sample. These five galaxies also have [OIII]$\lambda$5007 equivalent width about 200$-$300 \AA\ in their SDSS spectra. The \lya\ emission in LARS01 shows an offset from the UV emission, and is very similar to the four cases with \lya-UV offsets in our sample. The \lya\ emission in LARS05 shows partial central absorption and is very similar to GP1429$+$0643, the double-horned case in our sample. The 20\% Petrosian radius of \lya\ emission of these five galaxies are $2.3-3.6$ times larger than the 20\% Petrosian radius of \Ha\ emission (Hayes et al. 2014), which are very similar to the \lya/UV FWHM ratios in our sample.

\begin{figure}[ht]
\centering
  \includegraphics[width=0.5\textwidth]{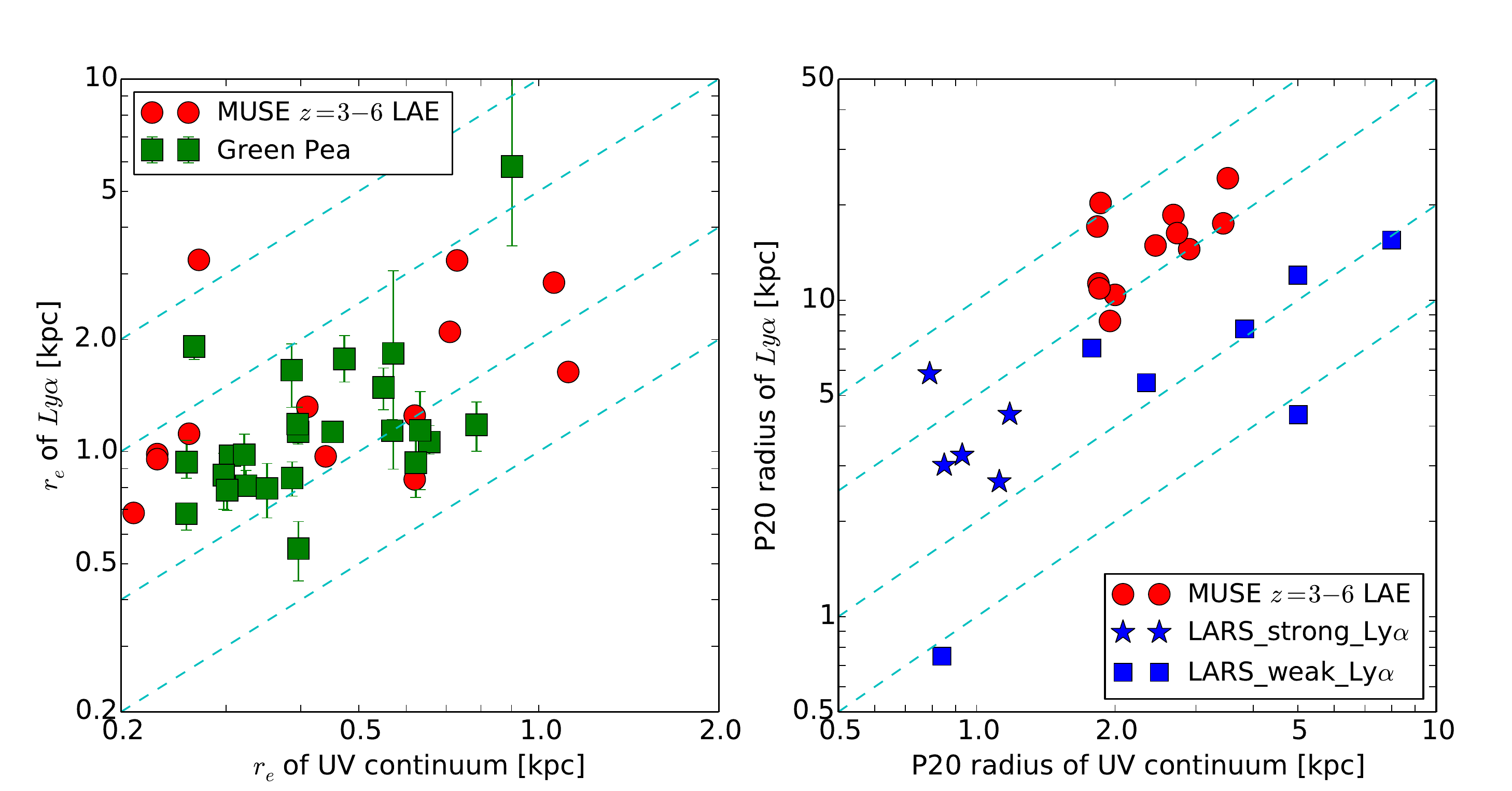}
  \caption{Left: Comparison of the \lya\ to UV scale length of Green Peas (green squares) and MUSE $z=3-6$ LAEs sample (red dots) (Wisotzki et al. 2016). The \lya\ scale lengths of MUSE LAEs are measured from the radial profile in Wisotzki et al. (2016) using the same method as Green Peas.  Right: Comparison of the \lya\ to UV Petrosian 20\% (P20) radius of LARS $\sim0$ galaxies and MUSE LAEs. Notice that the Petrosian 20\% radius of MUSE sample is measured from the best fit model of radial profile. The dashed cyan lines show constant ratios of 1:1, 2:1, 5:1, and 10:1.
}
\end{figure}

Two studies measure \lya\ sizes of 5 high-$\it{z}$ LAEs with high resolution HST narrow-band imaging (Bond et al. 2010; Finkelstein et al. 2011). Bond et al. (2010) suggested \lya\ sizes are compact and similar to UV emission; but Finkelstein et al. (2011) suggested the half light radius of \lya\ appears $\sim$1.6 times larger than the half light radius of UV continuum. These narrow band HST images of high-$z$ LAEs are very hard to get and have low S/N ratios, thus the \lya\ and UV sizes of low-$z$ LAEs are valuable. 
Since Green Peas are analogs of high-$\it{z}$ LAEs, our results suggest that most high-$\it{z}$ LAEs likely have larger \lya\ sizes than UV sizes. The extended \lya\ emission probably indicates gas outflows around galaxies illuminated by \lya\ light. 

One interesting question regards the redshift evolution of \lya\ sizes of LAEs. Recently, Wisotzki et al. (2016) measured \lya\ radial profiles of a sample of LAEs at $z=3-6$ from VLT/MUSE data and found that in 12 LAEs with both \lya\ and UV continuum sizes, the \lya\ light is considerably more extended than the UV continuum light. Here we compare the sizes of Green Peas and the MUSE LAEs. Using the \lya\ radial profiles in Wisotzki et al. (2016), we measured the deconvolved \lya\ scale radius $r_{e}$ assuming an intrinsic exponential profiles, so that the methods are same when measuring the $r_{e}$ of MUSE LAES and Green Peas. As shown in figure 5, the \lya\ to UV sizes ratios of Green Peas and MUSE LAEs are very similar. (Notice that some MUSE LAEs have extended \lya\ halos far beyond the scale radius. But for Green Peas, we don't have robust data to characterize the \lya\ emission beyond a few Kpcs.)

In the right panel of figure 5, we compare the Petrosian 20\% radius ($R_{P20}$) of MUSE LAEs (Table 2 in Wisotzki et al. (2016)) to that of the LARS sample (Table 1 in Hayes et al. (2013)). Compared to the five strong \lya\ emitters in LARS sample (marked by stars, LARS01, 02, 05, 07, and 14), the MUSE LAEs only have about 2 times larger ratios of $R_{P20}(Ly\alpha)$ to $R_{P20}(UV)$. One caveat of the comparison is that the Petrosian radius of MUSE sample is measured from the best fit {\it model} of radial profile, instead of the observed data, which is different from the method used in LARS sample. This might be the reason that the $R_{P20}(UV)$ of MUSE LAEs are  about $2-4$ kpc, about a factor of three larger than the $R_{P20}(UV)$ of the five LARS \lya\ emitters.

Based on our rough comparison of Green Peas and MUSE LAEs, the scale lengths of \lya\ and UV continuum have small evolution with redshift. This is not surprising considering that Green Peas and high-$z$ LAEs have very similar galactic properties such as stellar mass, star formation rate, and starburst age. The starburst in Green Peas and LAEs can drive gas outflows to the outer part of galaxies, and the gas outflows can scatter the \lya\ light and make the extended \lya\ emission.

\begin{figure}[ht]
\centering
  \includegraphics[width=0.5\textwidth]{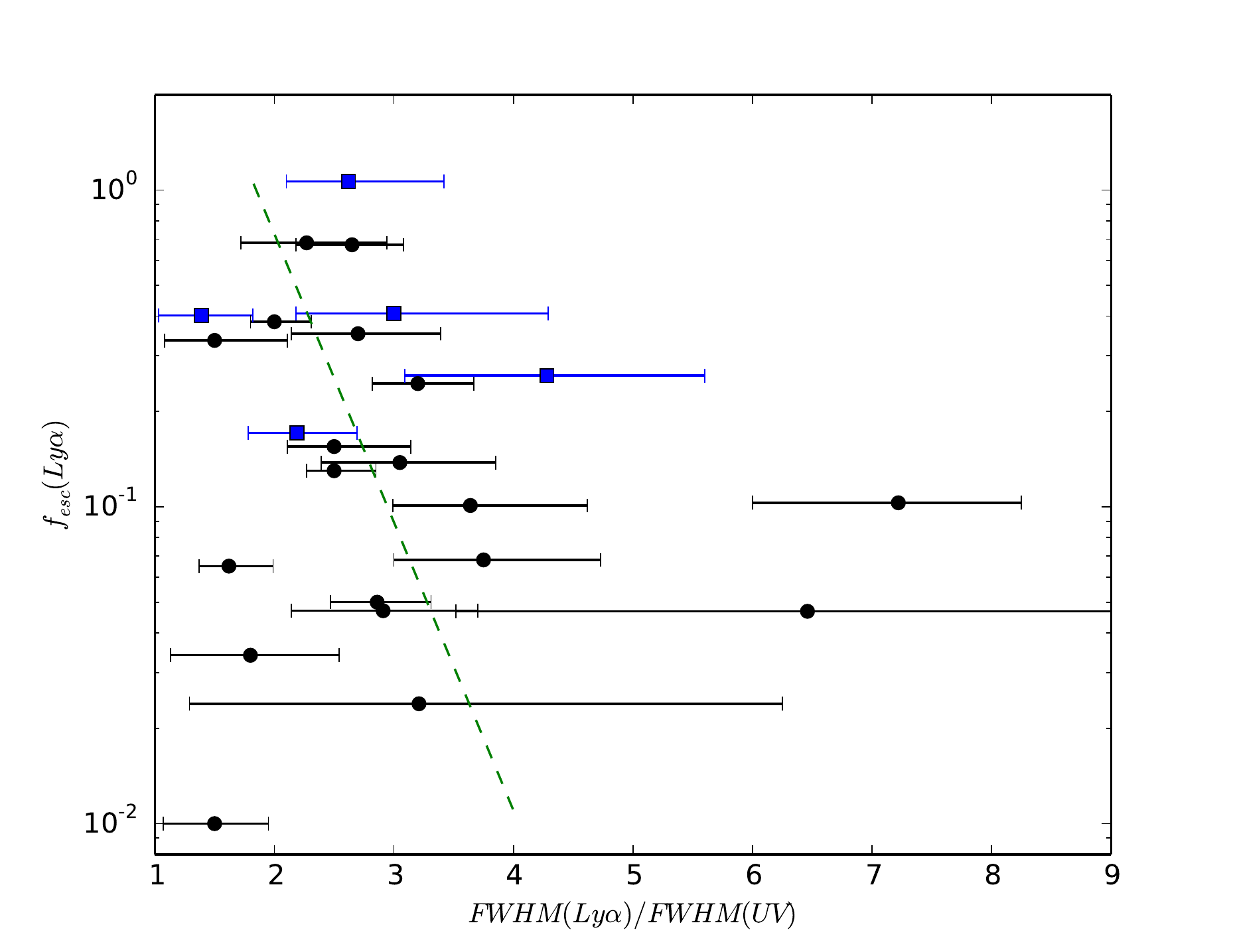}
  \caption{The relation between \fesc\ and the size ratio $FWHM(Ly\alpha)/FWHM(UV)$ (column (7) of Table 2). The blue square shows the five LyC leakers in this sample. The dashed green line shows a linear fit to the points with \fesc$>0.1$. The Spearman correlation coefficient for the points with with \fesc$>0.1$ is r=-0.41 with null probability=0.11.}
\end{figure}

\subsection{Implication for \lya\ and LyC Escape} 
Our results indicate \lya\ have larger sizes than the UV continuum. Since \lya\ is a resonant line, our results suggest most \lya\ photons escape out of galaxy through many resonant scatterings in the low HI column density gas in Green Peas. If there are fewer scatterings in the \lya\ escape process, the \lya\ escape fraction would be higher and the \lya\ emission would be more compact. So there may be an anti-correlation between \fesc\ and the size of \lya\ light. In figure 6, we show the relation between \fesc\ and the size ratio $FWHM(Ly\alpha)/FWHM(UV)$ (column (7) of Table 2). The scatters is large, but it shows a weak trend for objects with \fesc\ $\gtrsim 0.1$, indicating that LAEs with higher \fesc\ have {\it more compact} \lya\ morphology. 

In figure 6, we also mark out the five LyC leakers with blue squares. These LyC leakers have similar \lya\ to UV continuum size ratio to the other Green Peas. We note that the other Green Peas could be unknown LyC leakers, as their current UV spectra ranges don't cover the LyC emission. The LyC leakers have 1.4 to 4.3 times larger \lya\ sizes than the UV continuum sizes, so most HI gas, which scattered \lya\ emission, is unlikely to be transparent to the LyC emission. Therefore the LyC emission of these LyC leakers are probably escape through ionized holes in the interstellar medium.

\section{Conclusion}
We have investigated the \lya\ and UV sizes of Green Pea galaxies using their HST-COS 2D spectra. Our main results are as follows. 

\begin{enumerate}

\item We compared \lya\ and UV sizes from the 2D spectra and 1D spatial profiles and found that most Green Peas show more extended \lya\ emission than the UV continuum. We also measured the deconvolved FWHM of the spatial profiles as their \lya\ and UV sizes. The \lya\ sizes in most Green Peas of this sample are about 2 to 4 times larger than their UV continuum sizes.  We also found the five LyC leakers in our sample have larger \lya\ sizes than UV continuum sizes by 1.4 to 4.3 times. 

\item In eight Green Peas, we compared the spatial profiles of \lya\ photons at blueshifted and redshifted velocities, and found the blue wing of the \lya\ line has a larger spatial extent than the red wing in four Green Peas with comparatively weak blue \lya\ line wings. 

\item Since Green Peas are analogs of high-$\it{z}$ LAEs, our results suggest that most high-$\it{z}$ LAEs likely have larger \lya\ sizes than UV sizes. We also show that Green Peas and MUSE $z=3-6$ LAEs sample have similar \lya\ to UV continuum size ratios.

\item We compared \lya\ escape fraction with the size ratio $FWHM(Ly\alpha)/FWHM(UV)$ and found that for those Green Peas with \fesc\ $>10\%$, objects with higher \fesc\ tend to have {\it more compact} \lya\ morphology. 

\end{enumerate}

\acknowledgments
The imaging and spectroscopy data are based on observations with the NASA / ESA Hubble Space Telescope, obtained at the Space Telescope Science Institute, which is operated by the Association of Universities for Research in Astronomy (AURA), Inc., under NASA contract NAS 5-26555. Some of the data presented in this paper were obtained from the Mikulski Archive for Space Telescopes (MAST). STScI is operated by the Association of Universities for Research in Astronomy, Inc., under NASA contract NAS5-26555. Support for MAST for non-HST data is provided by the NASA Office of Space Science via grant NNX09AF08G and by other grants and contracts. H.Y. acknowledges support from China Scholarship Council. H.Y. and J.X.W. thanks supports from NSFC 11421303, CAS Frontier Science Key Research Program (QYZDJ-SSW-SLH006), and the Strategic Priority Research Program ``The Emergence of Cosmological Structures" of the Chinese Academy of Sciences (grant No. XDB09000000). Partial support for this work was provided by NSF grant AST-1518057.

\end{document}